\documentclass[aps,prl,reprint,longbibliography,superscriptaddress]{revtex4-1}
 \usepackage{amsmath,bm}
 \usepackage{mathrsfs}
 \usepackage{amsfonts}
 \usepackage{graphicx}
 \usepackage{setspace} 
 \usepackage{graphicx}
 \usepackage{epstopdf}
 \usepackage{dcolumn}
 \usepackage{amsmath}
 \usepackage{epsfig}
 \usepackage{indentfirst}
 \usepackage{psfrag}
 \usepackage{subfigure}
 \usepackage{amssymb}
 \usepackage{color}
 \usepackage{xcolor}
 \usepackage{siunitx}
 
 \usepackage{graphicx}
 \usepackage{dcolumn}
 \usepackage{bm}
 \usepackage{natbib}
\usepackage{physics}
\usepackage{natbib}
\usepackage[backref=none,bookmarksnumbered=true,bookmarks=true,bookmarksopen=true,colorlinks=true,
citecolor=blue,linkcolor=blue,anchorcolor=green,urlcolor=blue,unicode=false]{hyperref}

\usepackage{ulem}[normalem] 

\def\q0{$Q_0$}
\def\didv{$\mathrm{d}I/\mathrm{d}V$}

\newcommand{\reff}{$r_{\text{eff}}$}

\makeatletter

\newcommand\colorsout[1]{\bgroup \markoverwith{\textcolor{#1}{\rule[0.5ex]{2pt}{0.4pt}}}\ULon}

\AtBeginDocument{%
    \newwrite\bibnotes
    \def\bibnotesext{Notes.bib}
    \immediate\openout\bibnotes=\jobname\bibnotesext
    \immediate\write\bibnotes{@CONTROL{REVTEX41Control}}
    \immediate\write\bibnotes{@CONTROL{%
    apsrev41Control,author="08",editor="1",pages="1",title="0",year="1"}}
     \if@filesw
     \immediate\write\@auxout{\string\citation{apsrev41Control}}%
    \fi
}%

\makeatother
\begin{document}

\title{Local control of parity and charge in nanoscale superconducting lead islands}

\author{Stefano Trivini}
\thanks{These authors contributed equally}
  \affiliation{CIC nanoGUNE-BRTA, 20018 Donostia-San Sebasti\'an, Spain}

\author{Jon Ortuzar}
\thanks{These authors contributed equally}
  \affiliation{CIC nanoGUNE-BRTA, 20018 Donostia-San Sebasti\'an, Spain}

\author{Katerina Vaxevani}
  \affiliation{CIC nanoGUNE-BRTA, 20018 Donostia-San Sebasti\'an, Spain}

\author{Beatriz Viña-Bausá}
  \affiliation{CIC nanoGUNE-BRTA, 20018 Donostia-San Sebasti\'an, Spain}
 
\author{F. Sebastian Bergeret}
 \affiliation{Centro de F\'isica de Materiales (CFM-MPC) Centro Mixto CSIC-UPV/EHU, E-20018 Donostia-San Sebasti\'an,  Spain}
\affiliation{Donostia International Physics Center (DIPC), 20018 Donostia-San Sebastian, Spain}

\author{Jose Ignacio Pascual}
  \affiliation{CIC nanoGUNE-BRTA, 20018 Donostia-San Sebasti\'an, Spain}
\affiliation{Ikerbasque, Basque Foundation for Science, 48013 Bilbao, Spain}
\date{\today}
\begin{abstract}


Small superconducting islands can exhibit charge quantization, where Coulomb interactions compete with Cooper pairing. Using scanning tunneling spectroscopy, we probe this interplay by measuring the charging energy ($E_C$) and the pairing energy ($\Delta$) of individual nano-islands. Below a critical island size, where $E_C>\Delta$, we observe a crossover between even and odd parity ground states. By applying controlled voltage pulses, we continuously tune the island’s electrostatic potential and map the full charge-parity landscape. These results demonstrate tunable superconducting ground states, offering a potential platform for qubit design and control.

\end{abstract}
\date{\today}
\maketitle

Single-electron tunneling through a metallic nanoparticle capacitively coupled to two electrodes involves a Coulomb charging energy, $E_C$, determined by the capacitance of the junctions. In sufficiently small systems, $E_C$ can exceed the thermal energy, making it experimentally measurable~\cite{anderson1959,efros1975,likharev1988,grabert1992,hanna1991,vlaic2017}. This energy barrier give rise to sequential electron  tunneling and underlies the operation of Single-Electron Transistors (SETs)~\cite{grabert1992,altshuler1991}. 
In contrast to the repulsive nature of Coulomb interactions, superconducting correlations introduce an effective attractive interaction, characterized by the superconducting gap $\Delta$, which favors the formation of a Cooper-pair ground state~\cite{ingold1992,averin1986,vanHeck2016,altshuler1991,dittrich1998}. When such correlations are present, as in a superconducting SET, the competition between $E_C$ and $\Delta$ fundamentally modifies the standard SET response to a gate voltage, leading to a transition from single-electron to a two-electron periodic behavior~\cite{averin1986,Matveev1993,matveev1994}.

A particularly rich regime emerges when the charging energy $E_C$ and the superconducting gap $\Delta$ become comparable and compete in setting the ground-state parity. On one hand, pairing correlations weaken as the mean level spacing increases~\cite{vlaic2017,vondelft1996}; on the other, localized Cooper pairs remain protected by a parity gap modulated by the combined effects of $E_C$ and $\Delta$~\cite{Sacepe2020}. However, since both mechanisms lead to a gap in the density of states, disentangling their respective roles becomes experimentally challenging. Resolving this competition is essential for understanding the breakdown of superconductivity~\cite{Sacepe2020} and the emergence of insulating phases in nanoscale systems~\cite{ganguli2022,Carbillet2020}.

\begin{figure}[b]
        \begin{center}
			\includegraphics[width=0.99\columnwidth]{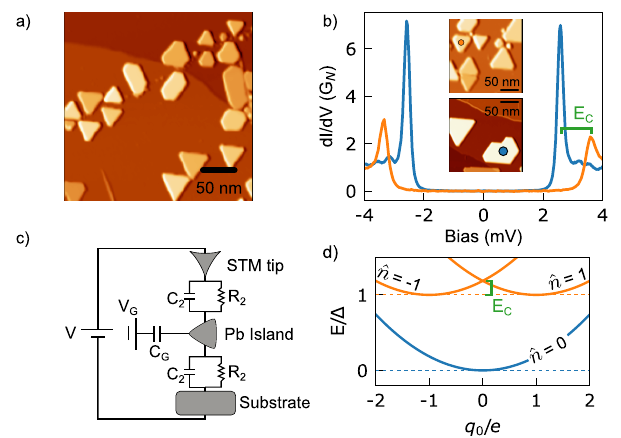}    
        \end{center}
			\caption{
            a) STM image of Pb islands on graphene. 
            (b) STS on a large (\reff$ = 39.2$~nm, blue) and small (\reff$ = 21.7$~nm, orange) island marked in the insets. A finite $E_C$ causes a larger gap in the smaller.    
            c) DBTJ circuit modeling the tip-island-sample geometry.
            d) Eigenstates of the DBTJ Hamiltonian, dashed lines for $E_C=0$, in blue the even states and orange the odd states.  STS @ R = 10 M$\Omega$; STM @ V = 400~mV, I = 20~pA.}
		\label{fig1}
\end{figure}

Here we demonstrate full electrostatic control over the parity of the superconducting ground state in individual Pb islands on graphene. Using scanning tunneling spectroscopy (STS), we access the regime where pairing and charging energies are comparable, and show that even-parity superconducting and odd-parity states can coexist and be reversibly tuned via gating potentials. The superconducting gap increases with decreasing island size, signaling the onset of Coulomb blockade. By tracking the gap evolution under magnetic field, we disentangle the contributions of pairing ($\Delta$) and charging energy ($E_C$), and identify a critical radius below which $E_C > \Delta$. In this regime, odd-parity ground states emerge when an excess charge of $\pm e$ is added to the island. We induce such transitions by applying voltage pulses with the STM tip, effectively gating the islands potential as in three-terminal geometries. These odd-parity states exhibit a characteristic magnetic-field-induced reopening of the gap, enabling their identification. The demonstrated control over $\Delta$, $E_C$, and the local potential opens new routes to engineer parity-based superconducting devices and probe the breakdown of Cooper pairing at the nanoscale.

\begin{figure}[t]
        \begin{center}
			\includegraphics[width=0.99\columnwidth]{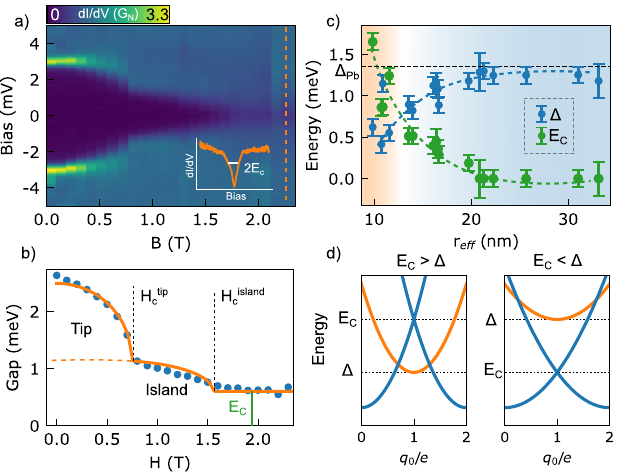}    
        \end{center}
			\caption{a) Evolution of the spectral gap with an out-of-plane magnetic field on a $r_{\text{eff}} = 15.9$~nm island. The dashed line marks the \didv\ spectrum of the bare Coulomb gap in the inset. 
            b) Gap values (blue dots) extracted by DBTJ fits to panel a); the orange curve results from the model in SM \cite{SI} that captures the sequential closing of the tip and sample gaps, yielding $E_C$ at high fields. 
            c) Extracted values of $\Delta$ and $E_C$ as a function of island size, with crossing at $r_{\text{eff}} \sim 12$~nm. 
            d) DBTJ eigenstates for varying $\Delta$ and $E_C$ at $q_0 \sim e$, with odd (even) ground state for $E_C > \Delta$ ($E_C < \Delta$).}
		\label{fig2}
\end{figure}



\textit{Description of the system.---}
We used graphene grown on a SiC single crystal \cite{badami1965,novoselov2004a}) as a substrate, and deposited on top a monolayer amount of lead at room temperature and under ultra-high vacuum. Lead forms flat, three-fold symmetric islands \cite{cortes-delrio2021,cortes-delrio2024,trivini2025} [Fig.~\ref{fig1}(a)] with effective radii $r_{\text{eff}}=(\text{area}/\pi)^{1/2}$ from 5~nm to more than 40~nm, and heights from 3 to 15~nm.

At the base temperature of our STM (1.2 K \cite{Specs}), islands with $r_{\text{eff}}>30$nm show a superconducting ground state similar to bulk lead. Differential conductance (\didv) spectra on these islands [e.g., blue curve in Fig.\ref{fig1}(b)] display a BCS-like gap with symmetric coherence peaks. Using Pb-coated superconducting tips with $\Delta_t\sim 1.3$ meV, the coherence peaks appear at $\Delta + \Delta_t \sim 2.6$ meV, resulting in a $\Delta\approx1.3$ meV for the island, close to the bulk value of lead ($\Delta_{\text{Pb}}=1.35$~mV). In contrast, smaller islands exhibit significantly larger spectral gaps, as shown in Fig.\ref{fig1}(b) for a $r_{\text{eff}}\sim 21.7$~nm island, with a gap of $\sim3.5$~meV. The larger gap is consistent with the existence of a charging energy $E_C$ due to Coulomb effects \cite{brun2012,wang2023b}.

To account for Coulomb blockade effects in the electronic transport through small superconducting islands, we use the double-barrier tunneling junction (DBTJ) model sketched in Fig.~\ref{fig1}(c). The Pb islands are capacitively coupled to the superconducting tip and the proximitized graphene sample \cite{cortes-delrio2024,trivini2025}, and their Hamiltonian contains the two main energy scales $E_C$ and $\Delta$~\cite{altshuler1991,dittrich1998,ingold1992,Devoret1990}:  
\begin{align}
    \hat H &= \hat H_S + \hat H_C \label{hamiltonian} \\  
    \hat H_S &= \dfrac{1 - (-1)^{\hat n}}{2} \Delta \label{Hs} \\
    \hat H_C &= \dfrac{(\hat n - q_0/e)^2}{2C} e^2 = E_C (\hat n - q_0/e)^2 \label{Hc}
\end{align}

where $\hat{H}_{S}$ accounts for the superconducting pairing, $\hat{H}_{C}$ considers the effect of charging the island and $C=C_1 + C_2$, the sum of tip-island and island-sample junctions capacitances. The full DBTJ model includes the tunneling Hamiltonian from the island to the superconducting electrodes (see Supplemental Material (SM) for additional details \cite{SI}). The superconducting pairing term $\hat{H}_S$ adds the energy cost of a quasiparticle excitation $\Delta$ only for an odd number of electrons in the island \cite{Matveev1993,matveev1994,averin1992,Eiles1993c,Hergenrother1994b}. In STM $C_1<<C_2$, \cite{ast2016a,senkpiel2020,serrier-garcia2013}, therefore the charging energy $E_C = e^2/2C_2$ depends mainly on the island-substrate capacitance. The magnitude of $E_C$ is generally negligible for large islands, but becomes sizable as $C_2$ decreases for small islands. In this case, changes of the integer charge in the island $\hat n$ upon electron transfer have an energy cost $E_C$. This value can be tuned by an electrostatic gate potential $V_G$, represented in eq.~\ref{Hc} as a residual (fractional) charge $q_0=V_GC_G$~\cite{ingold1992}. 

\begin{figure*}[t]
        \begin{center}
			\includegraphics[width=1\textwidth]{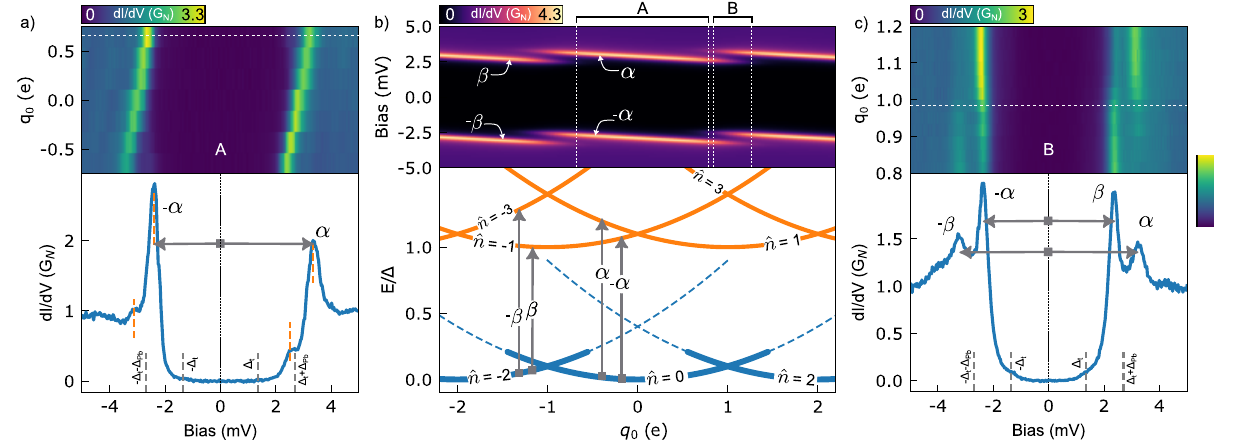}    
        \end{center}
			\caption{a) (bottom) Differential conductance on a Pb island with $r_{\text{eff}}= 26.8$~nm and $q_0\sim 0.75e$. We label $\alpha$ the dominant resonances occurring at asymmetric energy positions. (top) Excess charge dependence map of the island, obtained by consecutive bias pulsing. Each \didv\ spectrum is measured after a bias pulse between 1 and 2V.           
            b) (bottom) Eigenstates of the DBTJ Hamiltonian depending on the electron number $\hat{n}$ and on the excess charge $q_0$. Around $q_0=0$, we indicate with vertical arrows the transitions $\pm\alpha$ visible in the \didv\ of panel a). Close to $q_0=e$, the ground state population is a mixture of $\hat{n}=0,2$ and the possible transitions are $\pm\alpha$ and $\pm\beta$. (top) Calculated \didv\ with the DBTJ model as a function of the excess charge, $q_0$, with parameters: $\Delta=\Delta_t=1.35$ mV,$\Gamma=\Gamma_t=80$~uV, $C_1\approx0, C_2 = 480$~aF, $R_1=1000*R_2$, T = 1.3~K.  Two regions are highlighted to compare with the corresponding experimental maps. 
            c) (bottom) Differential conductance on a Pb island with \reff= 15~nm at $q_0\sim e$, showing symmetric excitations labeled $\alpha$ and $\beta$. (top)  Excess charge dependence map of the island around the point $q_0\sim e$ acquired similarly to the one in panel d).} 
   
		\label{fig3}
\end{figure*}

The eigenvalues of Eq.~\eqref{hamiltonian} in Fig.~\ref{fig1}(d) describe parabolic bands $E_n(q_0)$ shifted in $q_0$ by one elementary charge $e$. Superconducting pairing raises the parabolas for odd charge states (orange) by $\Delta$ relative to even ones (blue). At $q_0$ = 0, exciting a quasiparticle in the island requires both the pairing energy $\Delta$ and the charging energy $E_C$, so the spectral gap reflects their combined effect~\cite{ganguli2022}. As both depend on island size, separating their contributions is nontrivial.

We address this by applying an external magnetic field $H$ and monitoring the evolution of the superconducting gap. As shown in Fig.~\ref{fig2}(a) for an island with effective radius $r_{\text{eff}}=15.9$~nm, both the tip and island superconducting gaps gradually close with increasing $B$, while $E_C$ remains field-independent. The extracted gap and the theoretical fit appear in Fig.~\ref{fig2}(b) [details of the fit in SM \cite{SI}]. Starting from a zero-field gap of 2.6 mV, the tip gap ($\Delta_t$) closes first at a critical field of $H_c^{\text{tip}} \sim 0.8$ T (verified by measurements on bulk-like islands), followed by the island gap ($\Delta$) at $H_c^{\text{island}} \sim 1.5$ T (details in SM \cite{SI}). 
Thus, the remaining gap observed at higher fields [inset of Fig.~\ref{fig2}(a)] corresponds exclusively to the charging energy $E_C$ \cite{yuan2020}.

To reveal the size dependence of $E_C$ and $\Delta$, we applied this procedure to 18 islands with $r_{\text{eff}}$ from 10~nm to 35~nm. As shown in Fig.~\ref{fig2}(c), $E_C$ is negligible for large islands but increases significantly with decreasing island size. This dependence originates from the capacitance of the island-graphene interface, specifically  C$_2$ $\propto r_{\text{eff}}^2$. 

In contrast, the superconducting gap ($\Delta$) of the islands decreases as the island size shrinks, reflecting weakened superconducting correlations due to the increasing mean level separation ($\delta$) arising from confinement \cite{wang2023b} [more details about $\delta$ in SM \cite{SI}]. The plot in Fig.~\ref{fig2}(c) outline a transition to a regime where the charging energy exceeds $\Delta$ for island sizes below $r_c\sim12$~nm. In this regime, ground states with odd electron number (one extra quasiparticle) become accessible for an excess charge close to $q_0=e$ in the island, as depicted in the DBTJ model of Fig.~\ref{fig2}(d)~\cite{averin1992,grabert1992}.


\textit{Locally gating the Pb islands with voltage pulses.---} 
In contrast to typical particle-hole symmetric spectra of superconductors,  islands with effective radius $r_{\text{eff}}<25$~nm frequently exhibit coherence \didv\ peaks at bias-asymmetric positions and intensities [e.g. as in the spectrum of Fig.~\ref{fig3}(a)]. Such asymmetric spectra is originated by a finite electrostatic gate potential $V_G$,  which enters as a non-zero fractional residual charge $q_0$ in the DBTJ model of eq.~\ref{Hc} \cite{hanna1991}. As shown in Fig. \ref{fig3}(b), in the presence of Coulomb effects, deviations from the symmetric case ($q_0 = 0$) lead to unequal energy costs for adding ($\Delta\hat{n} = +1$, $\alpha$) or removing ($\Delta\hat{n} = -1$, $-\alpha$) an electron. This imbalance shifts the coherence peaks depending on the bias polarity, revealing built-in electrostatic potentials at the island–graphene interface.

Controllably gating surface-supported objects has long been a challenge in STM experiments \cite{brar2011}. In the Pb/graphene system, we could controllably modify the the excess charge  $q_0$ of small  islands by applying controlled bias pulses with the STM tip \cite{hong2013}. As shown in  Fig. \ref{fig3}(a) (upper panel), repeated bias pulses at increasing voltage amplitude, spanning in the range from $3$~V to $-3$V, induce reproducible changes in the spectral shape, associated to gradual variations of the residual charge $q_0$ by more than one electron charge (see SM for details \cite{SI}). 
Using this gating method we acquired gate-dependent spectral maps covering different regions of accumulated charge in the islands. 
The spectral map in Fig. \ref{fig3}(a) shows  coherence peaks of an small island shifting markedly after pulses,  with unmodified superconducting gap width. This proves that the island's spectral asymmetry arises from a net local gating $q_0$, which is modified after each voltage pulse in the window $\sim\pm e$.



The spectral map in Fig.~\ref{fig3}(c), showing four \didv\ peaks with varying intensity and position, indicates that the island is near an odd value of residual charge. This behavior is well captured by the DBTJ model in Fig.~\ref{fig3}(b), which predicts a degeneracy of the $\hat{n} = 0$ and $\hat{n} = -2$ ground states at $q_0 = -e$, as well as particle-hole symmetric excitations to the $\hat{n} = -1$  and the $\hat{n} = -3$ states. From the $\hat{n} = -2$ ground state, an additional electron (hole) excitation to $\hat{n} = -1$ ($\hat{n} = -3$) becomes now accessible, accounting for a second pair of excitation peaks ($\pm\beta$) close to $q_0=e$. The simultaneous observation of $\alpha$ and $\beta$ coherence peaks with gradually changing amplitude around the crossing of the degeneracy point is attributed to thermal population of both even-parity ground states (see SM~\cite{SI}). This agreement enables a direct and accurate determination of odd residual charge states in  islands and confirm the persistence of their paired ground states for sizes above the critical effective radius.





 \begin{figure}[t]
        \begin{center}
			\includegraphics[width=0.48\textwidth]{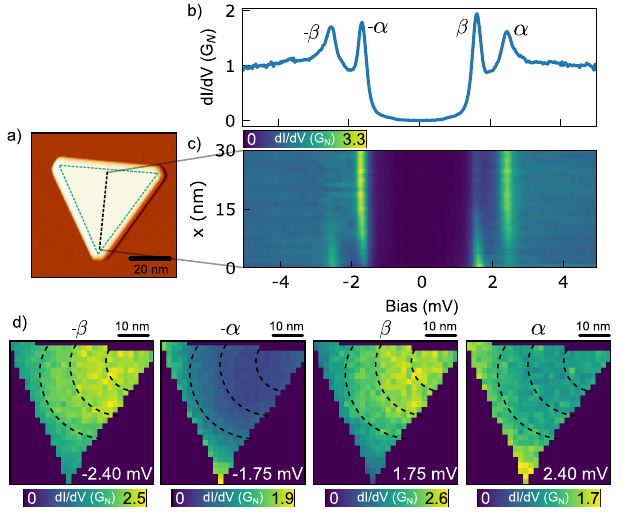}    
        \end{center}
			\caption{a) Topographic image of an island with \reff= 21.2~nm. b) Measured \didv\ spectrum on the island with the symmetric resonance labeled $\pm\alpha$ and $\pm\beta$ indicating $q_0\sim e$. c) Spatial dependence of the LDOS along the black dashed line in panel a). d) Constant energy cuts at $\pm\alpha$ and $\pm\beta$ of a spectral grid acquired over a triangular region on the island in panel a). The radial dashed line indicates the shape of the charge puddle induced in graphene, which modulates locally the excess charge. 
            }
		\label{fig4}
\end{figure}




\textit{Local mapping of the ground state charge.---} 
As it has been previously shown~\cite{cortes-delrio2024}, lead islands can be repositioned around the graphene surface through lateral manipulation with the STM tip. After their movement, we often observe changes or even reversals of spectral asymmetry (see SM~\cite{SI}), likely due to localized charges or built-in potential gradients in the substrate~\cite{hong2013,brar2011}. This suggests that the island’s residual charge is sensitive to variations of the electrostatic landscape of the surface and could be used as a sensor for point charges.

Figure~\ref{fig4} illustrates this concept for a $r_{\text{eff}} \sim 21.2$~nm island with ground state at the transition between $\hat{n}=0$ and $\hat{n}=-2$, as determined by its spectrum in Fig.~\ref{fig4}(b). Two \didv\ peaks at positive and negative bias are consistent with a residual charge of $q_0 \sim e$.  
However, as shown in the spectral profile in Fig.~\ref{fig4}(c), the intensity (and, weakly, the alignment) of the four peaks changes with the position along the center of the island.
This evolution resemble the spectral changes with $q_0$  shown in Fig.~\ref{fig3}(c), thus suggesting an inhomogeneous electrostatic potential at the island-graphene interface. 

To map the potential landscape, we measured the spatial distribution of the four spectral peaks $\pm \alpha$ and $\pm \beta$ across the island using a grid of STS measurements (within the dashed triangle in Fig.~\ref{fig4}(a)). The radial distribution of peak amplitudes shown in Fig.~\ref{fig4}(d) suggests the presence of a point charge near one of the island’s corners. 
The $\beta$ transitions, representing excitations from the ground state with a Cooper pair less ($-2e \rightarrow -e$ and $-2e \rightarrow -3e$), are strongest toward the upper right corner and fade toward the center. In contrast, 
$\pm \alpha$ peaks, corresponding to the $0 \rightarrow e$ and $0 \rightarrow -e$ transitions, are more intense near the left edge.
The predominance of the $q = -2e$ ground state ($\pm\beta$ excitations) in the right corner likely results from the localization of a negative point charge in this region. Such static charge puddles are common in graphene on insulating substrates and can be created, neutralized, or reversed by applying STM voltage pulses~\cite{brar2011}. Here, their field acts as a local gate, tuning the island’s residual charge.

\begin{figure}[b]
        \begin{center}
			\includegraphics[width=0.98\linewidth]{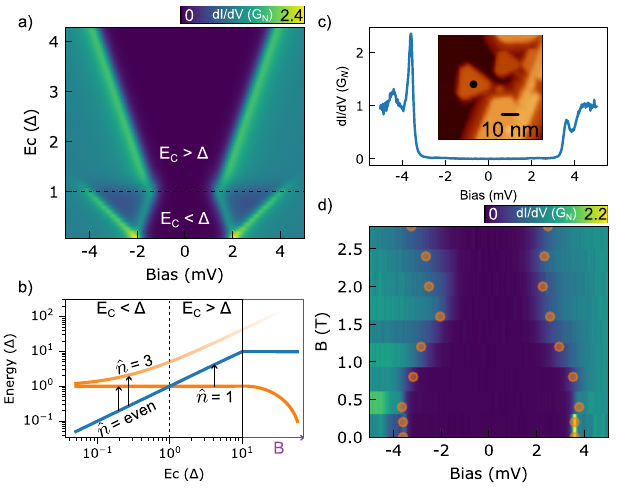}    
        \end{center}
			\caption{a) Calculated LDOS map varying the excess charge. The even parity ground state region with $E_C<\Delta$ is separated by a dashed line from the odd parity ground state region with $E_C>\Delta$. (params) 
            b) Ground and excited states increase the charging energy. In the right part of the scheme, we sketch the effect of the magnetic field, which, in an odd parity ground state, opens the energy gap. 
            c) Spectrum on a island with \reff= 9.5~nm, with gap $\sim3.8$~mV. 
            d) Magnetic field dependence of the LDOS measured on the same island. The orange dots are fits of the gap magnitude, which initially decreases, due to the closing of $\Delta_t$ and increases at higher fields, indicating an odd parity ground state.} 
   
		\label{fig5}
\end{figure}



The doubly-peaked gapped spectrum is a hallmark of the persistence of a superconducting (paired) ground state in small islands with electrostatic potentials near the $q_0 = e$ point. As shown in the DBTJ simulations in Figs.~\ref{fig5}(a,b), the separation between the two excitations increases with $E_C$, i.e., inversely with island size, and the outer peak disappears at the crossing $E_C = \Delta$. This marks the transition to an odd-parity ground state for $q_0 = e$, characterized by a single gap of width $2(E_C - \Delta)$. In this regime, the remanent electron pairing partially compensates for Coulomb correlations. A distinctive behavior of such odd-parity state is the evolution of its gap with magnetic field $H$ \cite{shen2018}. As superconductivity is suppressed, the gap widens until the bare Coulomb blockade gap emerges [right side of Fig.~\ref{fig5}(b)]. 


This peculiar  behavior is demonstrated in Figure~\ref{fig5}(c) and ~\ref{fig5}(d) for an island with \reff = 9.5~nm, smaller than the 12~nm crossover from Fig.~\ref{fig2}(c), and spectral gap of 3.8 mV, consistent with $E_C > \Delta$. 
This island could lie near either $q_0 = 0$ or $q_0 = e$, corresponding to an even or odd ground state, respectively, as shown in Fig.~\ref{fig2}(d).
Figure~\ref{fig5}(d) demonstrates that this island lies in fact in an odd charge ground state by plotting the evolution of the spectral gap with $B$. The gap closes initially with $B$ due to the suppression of the tip’s superconductivity, whereas the subsequent reopening at higher fields demonstrates the reappearance of the larger Coulomb blockade gap as superconductivity is quenched. This is the signature of an odd parity ground state stabilized on the island.

\textit{Conclusions.---}
In summary, we have shown that superconducting correlations compete with Coulomb blockade in lead islands with radius amounting to tens of nanometers. This interplay of interactions is detected as wider gaps in the density of states measured with scanning tunneling spectroscopy.   We used a magnetic field to disentangle the different roles of electronic pairing and Coulomb effects in the gap and found a critical effective radius amounting to 12~nm  below which Coulomb blockade dominates and even-odd parity ground states coexist and can be tuned with electrostatic gating.  In our work, we span a wide range of Coulomb blockade and superconducting pairing interactions controlled by the size of the particle, and tuned gating potential by applying voltage pulses. This allowed us to provide a convincing picture of the dominant role of Coulomb blockade in the destruction of superconductivity in small clusters. In combination with DBTJ model, we show that smaller clusters can stabilize odd-charge-parity superconducting ground states. The tunability of this versatile platform is ideal for exploring principles of fabricating  $\pi$-junction superconducting devices, with promising role in the realization of topological qubits~\cite{averin1992,albrecht2016,plugge2017,lutchyn2018}.   

\begin{acknowledgments}
We acknowledge financial support from Grants No. PID2019-107338RB-C61, No.  CEX2020-001038-M, No. PID2020-112811GB-I00, and Grant PID2020-114252GB-I00, funded by MCIN/AEI/ 10.13039/501100011033, from the Diputación Foral de Guipuzcoa, and from the European Union (EU) through the Horizon 2020 FET-Open projects SPRING (No. 863098),  and the European Regional Development Fund (ERDF).
J.O. acknowledges the scholarship   PRE\_2021\_1\_0350   from the Basque Government. B.V. acknowledges funding from the Spanish Ministerio de Universidades through the PhD scholarship No. FPU22/03675.
F.S.B. acknowledges the Spanish MCIN / AEI / 10.13039 / 501100011033 through projects PID2023-148225NB-C32 (SUNRISE) and TED2021-130292B-C42, and the EU Horizon Europe program [Grant Agreement No. 101130224 (JOSEPHINE)].
\end{acknowledgments}

\end{document}